\documentclass[aps,amsfonts,amsmath,amssymb,nofootinbib,twocolumn,superscriptaddress]{revtex4}

\usepackage{graphicx}
\usepackage{bm}
\usepackage{xcolor} 
\usepackage{soul}
\usepackage{todonotes}
\usepackage[normalem]{ulem}
\usepackage{multirow}
\usepackage[T1]{fontenc}

\usepackage{float}
\restylefloat{table}

\usepackage{natbib}
\setcitestyle{square}
\setcitestyle{comma}

\usepackage{hyperref}
\hypersetup{
     colorlinks = true,
     linkcolor = blue,
     anchorcolor = blue,
     citecolor = blue,
     filecolor = blue,
     urlcolor = blue
     }

\begin{document}
\title{Time-Domain Terahertz Spectroscopy in High Magnetic Fields}
\author{Andrey Baydin}\email{baydin@rice.edu}\affiliation{Department of  Electrical and Computer Engineering, Rice  University, Houston, TX 70005, USA}
\author{Takuma Makihara}\affiliation{Department of Physics and Astronomy, Rice University, Houston, Texas 77005, USA}
\author{Nicolas Marquez Peraca}\affiliation{Department of Physics and Astronomy, Rice University, Houston, Texas 77005, USA}
\author{Junichiro Kono}\email{kono@rice.edu}\affiliation{Department of  Electrical and Computer Engineering, Rice  University, Houston, TX 70005, USA}
\affiliation{Department of Physics and Astronomy, Rice University, Houston, Texas 77005, USA}
\affiliation{Department of Materials Science and NanoEngineering, Rice University, Houston, Texas 77005, USA}
\date{\today}

\begin{abstract}

There are a variety of elementary and collective terahertz-frequency excitations in condensed matter whose magnetic field dependence contains significant insight into the states and dynamics of the electrons involved. Often, determining the frequency, temperature, and magnetic field dependence of the optical conductivity tensor, especially in high magnetic fields, can clarify the microscopic physics behind complex many-body behaviors of solids. While there are advanced terahertz spectroscopy techniques as well as high magnetic field generation techniques available, a combination of the two has only been realized relatively recently.  Here, we review the current state of terahertz time-domain spectroscopy experiments in high magnetic fields. We start with an overview of time-domain terahertz detection schemes with a special focus on how they have been incorporated into optically accessible high-field magnets.  Advantages and disadvantages of different types of magnets in performing terahertz time-domain spectroscopy experiments are also discussed.  Finally, we highlight some of the new fascinating physical phenomena that have been revealed by terahertz time-domain spectroscopy in high magnetic fields.

\end{abstract}

\maketitle

\tableofcontents

\section{Introduction}

Terahertz (THz) time-domain spectroscopy (THz-TDS) has proven itself to be an invaluable tool for studying materials. Particularly, it is an excellent probe for various low-energy excitations in condensed matter systems, such as intraband transitions, superconducting gap excitations, phonons, magnons, and plasmons. One of the advantages of THz-TDS over traditional far-infrared spectroscopy based on Fourier-transform infrared spectrometers is its ability to measure the amplitude and phase of the transmitted or reflected THz electric field simultaneously, which allows one to determine the complex optical conductivity and refractive index of materials. There exist a number of good review articles about the basics of THz-TDS~\cite{NussOrenstein98THz,Schmuttenmaer04CR,Lee09Book,JepsenetAl11LPR,UlbrichtetAl11RMP, neu2018tutorial}, and thus, we will focus on THz-TDS combined with high magnetic fields.

Matter placed in a magnetic field provides a fascinating laboratory in which to study exotic quantum phenomena in a highly controllable manner. A magnetic field can control the states of electrons through spin and orbital quantization while inducing nonintuitive topological modifications and quantum interference of electron waves, especially at low temperatures.  These magnetic field-induced changes can drastically alter absorption, reflection, and emission spectra in the THz spectral range, which can be compared with many-body calculations to elucidate the origins of complex behaviors of materials.

The application of a high magnetic field allows one to enter a system-dependent \textit{high-field regime} where new states or phenomena are expected to arise.  For example, in an electron-hole system in a magnetic field, one enters the high-field regime when the cyclotron energy, $E_\text{c}$, becomes larger than the exciton binding energy, $E_\text{X}$.  When $\gamma = E_\text{c}/E_\text{X} \ll 1$, the magnetic field is a small perturbation, and the main visible effects are Zeeman splittings and diamagnetic shifts~\cite{CongetAl18InBook}. Exciting things begin to happen when one achieves the high-field regime ($\gamma >1$), where Landau quantization becomes important and quantum Hall physics kicks in. Most importantly, when $\gamma \gg 1$, one enters the \textit{high-field limit}, where ``hidden symmetry'' protects excitons against dissociation even at ultrahigh electron-hole pair densities~\cite{MacDonaldRezayi90PRB,DzyubenkoLozovik91JPA,ApalkonRashba91JETP,RashbaetAl00SSC}.

Another example is unconventional superconductors, such as high-$T_\text{c}$ cuprates, for which a high magnetic field is required to suppress superconductivity~\cite{proust2019remarkable,shi2020vortex} and study its normal state at low temperatures. Moreover, a high magnetic field can stabilize a magnetic field-induced superconductivity phase; for instance, UTe$_2$ has the highest magnetic field range identified for a re-entrant superconductor ($>$65\,T)~\cite{ran2019extreme}. In strongly correlated systems, a quantum critical point is usually hidden inside a superconducting phase, and thus, it is important to be able to destroy superconducting order to reveal the physics of quantum criticality. In two-dimensional (2D) materials, typically large effective masses require one to use high magnetic fields to study integer and fractional quantum Hall effects~\cite{dean2011multicomponent} and to enter the quantum limit state~\cite{moll2016magnetic}. Another important topic is Fermi surface topology studies through quantum oscillations in high magnetic fields, which occur only when Landau level spacings become larger than disorder or scattering rates.

To date, THz-TDS combined with high magnetic fields has enabled investigations of many-body physics of different types of modern materials, including quantum Hall systems, graphene and related 2D materials, topological insulators, and rare-earth orthoferrites, as detailed in Section~\ref{highlights}.  Below, after briefly describing the basics of THz-TDS (Section~\ref{THz-TDS}), we review the state-of-the-art experimental setups combining THz-TDS and magnetic fields (Section~\ref{THz-B}) and their ensuing results that show new exciting physics (Section~\ref{highlights}). 

\section{Terahertz Time-Domain Spectroscopy} \label{THz-TDS}

\subsection{Basics of THz-TDS}

\begin{figure*}
    \centering
    \includegraphics[width=\textwidth]{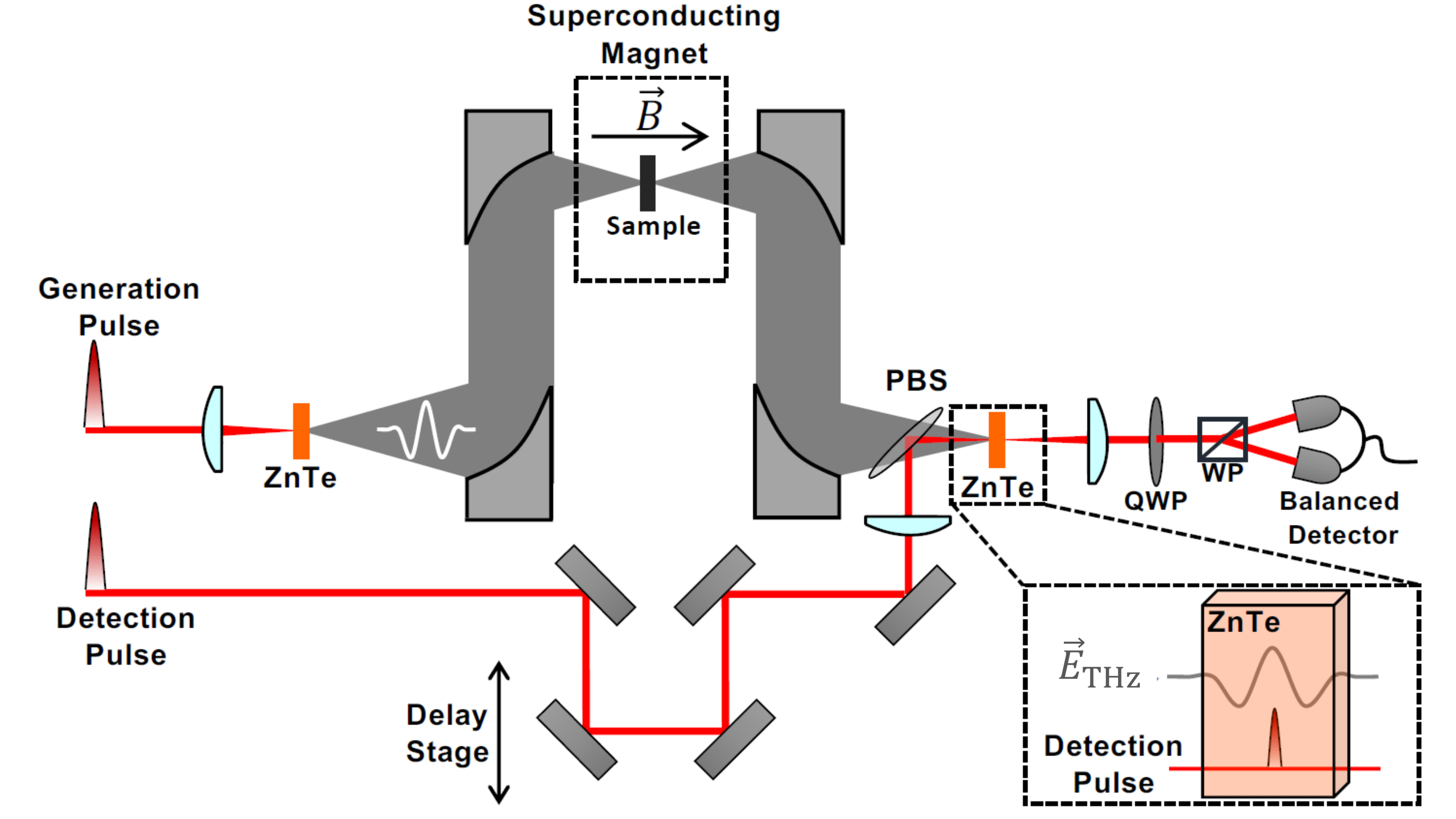}
    \caption{\textbf{Free-Space THz time-domain spectroscopy system combined with a superconducting magnet}. The generation and detection of THz radiation is achieved via optical rectification and the electro-optic effect, respectively, using ZnTe crystals. A pellicle beamsplitter (PBS) is used to combine the THz and near-infrared probe beams. A quarter-wave plate (QWP) and a Wollaston prism (WP) are commonly used to detect the polarization rotation of the probe beam. }
    \label{fig:SC_magnet}
\end{figure*}

A burst of electromagnetic radiation containing frequency components in the THz ($10^{12}$\,s$^{-1}$) range can be generated by exciting a crystal with femtosecond (fs) laser pulses.  The most common generation process employed is optical rectification, which is a second-order nonlinear optical process, while inducing a short real current pulse in a photoconductive antenna (PCA) is also often used. The choice of the generation process, and thus the THz emitter crystal, will define the bandwidth of the emitted THz radiation, which can range from $\sim$0.2~THz to $\sim$120~THz~\cite{neu2018tutorial}. Commonly employed emitters include ZnTe, LiNbO$_3$, and PCAs. Once a THz wave is generated, it needs to be guided through the magnet bore and eventually detected. The detection scheme used also determines the choice of the generation mechanism as the strength of the THz pulse needs to be taken into account for the best suited detection scheme, including single-shot detection approaches needed for THz-TDS with pulsed magnets. 

One of the most widely used methods for detecting THz radiation is based on free-space electro-optic (EO) sampling~\cite{wu1995free}. Based on a second-order nonlinear optical effect referred to as the Pockels effect, this method employs the modulations of the detection crystal's birefringence induced by the applied THz electric field. In free-space EO detection schemes, the THz electric field pulse is overlapped with an ultrashort probe pulse in the nonlinear optical detection crystal. By measuring the polarization rotation of the probe pulse induced by the THz electric field, one can determine the applied THz electric field strength. Some of the commonly used EO crystals are ZnTe~\cite{wu1996ultrafast,NahataetAl96APL}, GaP~\cite{WuZhang97APL}, and GaSe~\cite{HuberetAl00APL,LiuetAl04APL}, and they have different bandwidths.

Another popular method for detecting THz radiation uses a PCA. This method has been used since the 1980s~\cite{smith1988subpicosecond}. PCA-based THz radiation detectors use an ultrashort probe pulse to generate transient charge carriers in a semiconductor (i.e., a time-dependent photocurrent), which are then driven by the instantaneous THz electric field at a time determined by the position of the delay stage. Therefore, measuring the photocurrent as a function of time delay enables sampling of the THz electric field waveform in the time domain. 

A more recently developed methodology for THz radiation detection, referred to as the THz-air breakdown coherent detection (THz-ABCD) method, exploits the dependence of the efficiency of an optical probe pulse’s second-harmonic generation in air in the presence of an applied THz electric field to measure ultrabroadband THz pulses~\cite{LuetAl09JOSAB}. This method is advantageous as it circumvents the appearance of phonon modes that are present in solid state THz detectors.

\subsection{Temporally Sampling THz Electric Fields} \label{Sampling Section}
As mentioned in the previous section, detection of THz electric fields requires ultrashort laser pulses. These probe pulses are typically orders of magnitude shorter in duration than THz pulses. Here, we briefly review a few techniques for sampling the entire THz waveform in the time domain using such ultrashort laser pulses. As discussed in the subsequent sections, different techniques for temporally sampling the THz electric field are amenable to different types of magnets.

The most common method for temporally sampling a THz electric field uses a step-scan delay stage to adjust the time delay between the THz electric field and the probe pulse. In this technique, identical measurements must be repeated while systematically adjusting the delay between the probe pulse and the THz electric field until the entire THz waveform has been sampled. A particular advantage of using a delay stage is that measurements can be arranged with two mirrors or a retroreflector and a mechanical delay stage in a straightforward manner. Given the requirement for repeating identical measurements, using a delay stage is appropriate for measurements in a constant applied magnetic field. 

\subsubsection{Rapid-Scan Approaches}

\paragraph{ASOPS}
When THz-TDS is combined with a pulsed magnet, the entire THz waveform must be sampled rapidly. This precludes the use of a step-scan, mechanical delay stage. One of the methods to rapidly acquire the entire THz waveform in the time domain is asynchronous optical sampling (ASOPS)~\cite{ElzingaetAl87AO}. ASOPS requires two lasers, one used for THz generation and the other used for detection.  There is a finite offset in pulse repetition frequency, ($\Delta f$). As a result, the time delay between a generation pulse and a detection pulse is systematically delayed until the detection pulse has sampled the entire temporal region between two generation pulses in a time duration $1/\Delta f$. ASOPS was first applied to THz-TDS in 2005~\cite{janke2005asynchronous,yasui2005asynchronous} and has subsequently been improved to enable time delays of up to $1$~ns to be scanned in $100$~$\mu$s~\cite{BartelsetAl07RSI}. Not only does ASOPS enable rapid THz detection, it also eliminates certain challenges associated with using mechanical delay stages, such as probe beam misalignment as a result of mechanical motion. ASOPS has been demonstrated in THz TDS measurements in pulsed magnetic fields up to 31~T~\cite{spencer2016terahertz}.

\paragraph{ECOPS}
A method closely related to ASOPS for rapidly recording the entire THz waveform is electronically controlled optical sampling (ECOPS)~\cite{tauser2008electronically}. ECOPS also exploits two frequency-offset lasers, but with a time-varying frequency offset. ECOPS can be more desirable than ASOPS in many cases because it samples only a small time window following the generation pulse, enabling much faster acquisition. This is particularly relevant for THz-TDS, where the relevant physics being studied subsists for short times following the pump. For example, ASOPS using a $100$~MHz repetition rate laser would sample the full $10$-nanosecond time window between pulses~\cite{tauser2008electronically}, whereas THz-TDS is often concerned with picosecond timescales following the pump. In ECOPS, the frequency offset between the two lasers is modulated by modulating the cavity length of one laser. ECOPS has been demonstrated in THz-TDS measurements~\cite{KimetAl10OL,liu2010measurement} and for THz-TDS in pulsed magnetic fields up to 30~T~\cite{NoeetAl14AO}.

\paragraph{Rotating Delay Line}
Another type of technique for rapidly sampling the entire THz waveform to enable THz-TDS in pulsed magnetic fields uses a rotating delay line~\cite{MolteretAl10OE}. Here, the time-delay between the THz electric field and the detection probe pulse is rapidly modulated by reflecting the detection probe pulse off of an angled mirror mounted on a rotating element. Using this technique enabled 140~ps time-delays to be sampled at a repetition rate of 250~Hz in a 40~T pulsed magnetic field~\cite{MolteretAl10OE}.

\subsubsection{Single-Shot Approaches}

\paragraph{Echelons}
The best approach to the high-speed sampling of the entire THz waveform is single-shot detection, wherein a single ultrashort probe pulse samples the entire THz waveform. Recently, Teo \textit{et al}.~thoroughly reviewed the state-of-the-art single-shot THz detection techniques~\cite{teo2015invited}. They concluded that using transmissive echelons is robust and easily implementable, enabling single-shot detection with greater signal-to-noise ratios and faster acquisition times than a conventional delay stage. They also suggest that using reflective echelons~\cite{MinamietAl13APL}, as opposed to transmissive ones, would provide better frequency resolution and longer time windows.\\
\\
The use of echelons for time-resolved ultrafast spectroscopies was pioneered in the 1970s~\cite{topp1971absorption,topp1971dye}. In these measurements, a wide range of time delays between the pump and detection pulses was achieved by separating the probe pulse into time-delayed components using an echelon to alter the optical path length traveled by different portions. The first demonstration of single-shot THz-TDS using echelons employed two transmissive echelons, where the use of dual echelons enabled a longer time window to be observed~\cite{Kim07}. In echelon-based THz-TDS, EO sampling is typically used to map the THz electric field onto the polarization of the time-delayed components of the optical detection pulse. The time-resolved polarization of the probe can be quantitatively analyzed by imaging the echelon onto a camera. More recently, reflective echelons have been used to demonstrate single-shot THz-TDS~\cite{KatayamaetAl11JJAP,MinamietAl13APL}. Reflective echelons can be advantageous compared to transmissive ones because they eliminate complications arising from having an ultrashort probe pulse propagating through a delicate optical component, such as chirp or defects in the material. As a result, reflective echelons have been used to demonstrate single-shot THz-TDS in pulsed magnetic fields up to 30~T~\cite{NoeetAl16OE,MakiharaetAl20arXiv}.

\paragraph{Spectral Encoding}
Another technique for enabling single-shot THz-TDS is spectral encoding. In this technique, the different frequency components of a chirped optical probe pulse are used to sample the entire THz waveform~\cite{jiang1998electro,jiang1998single}. However, there exists a trade-off between time-resolution and the time window sampled by the chirped pulse, as longer time windows require mapping the THz electric field onto a narrower frequency spectrum, which becomes increasingly more challenging to resolve on a spectrometer. These limitations can be overcome using spectral interferometry~\cite{matlis2011single}, yet the sensitivity of this method to optical alignment~\cite{teo2015invited} should make implementing this single-shot detection technique challenging in large magnets.

In summary, rapid-scan and single-shot approaches both provide viable techniques for combining THz-TDS with time-dependent (pulsed) magnetic fields. Notably, there exist commercially available ASOPS and ECOPS platforms specifically designed for ultrafast THz-TDS~\cite{noe2014rapid}, whereas echelons suffer from requiring precisely machined components and difficulties associated with aligning non-traditional optics. Further, when combined with pulsed magnetic fields, rapid scanning techniques have demonstrated the acquisition of up to 100~THz waveforms for a single magnetic field pulse~\cite{spencer2016terahertz}, providing extremely fine magnetic field-dependent data. Nonetheless, single-shot techniques represent the ultimate limit of ultrafast THz sampling and overcome challenges arising from making repeated measurements, such as measurement-induced damage to the sample. Further, in the authors' experience, reflective echelon-based single-shot THz-TDS provides the best signal-to-noise ratios and thus merits our recommendation.

\section{Combining THz-TDS and High Magnetic Fields}\label{THz-B}

The magnets that have been used for high-field THz-TDS can be classified into three categories: (i)~DC superconducting magnets, (ii)~DC resistive magnets, and (iii)~pulsed magnets. Here, we briefly review the benefits and shortcomings of each magnet when used for THz time-domain magnetospectroscopy. We also discuss a few representative magnets for each category. 

\subsection{DC Superconducting Magnets}
For individuals seeking to build a high-field THz time-domain magnetospectrometer, superconducting magnets should be considered first. These magnets are advantageous compared to resistive magnets, particularly in university settings, because they demand significantly less electrical power to operate. As suggested by their name, superconducting magnets only require a bath of liquid helium to maintain a coil in its superconducting state. For THz-TDS, these magnets are also advantageous as they provide easy optical access to samples in static magnetic fields. THz radiation is challenging to manipulate over large distances, so magnets with easy optical addressability are crucial. Further, as mentioned in the previous section, static magnetic fields also allow for the simplest implementation of THz detection, namely, using a step-scan, mechanical delay stage. Static magnetic fields also allow for arbitrarily large numbers of THz waveforms to be averaged at little cost. However, the magnetic field produced by a typical laboratory superconducting magnet with good optical access is limited to roughly 10~T. Nonetheless, their ease of implementation and operation has made the use of superconducting magnets and delay stage-based THz electric field sampling the most popular form of THz-TDS in high magnetic fields~\cite{walecki1993terahertz,SomeNurmikko94APL,SomeNurmikko94PRB,SomeNurmikko96PRB,Crooker02RSI,WangetAl07OL,sumikura2007development,ikebe2008characterization,ScalarietAl12Science,george2012terahertz,wood2013chip,WuetAl16Science}.

\begin{figure}
    \centering
    \includegraphics[width = 0.45 \textwidth]{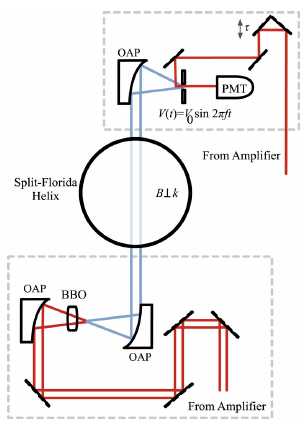}
    \caption{\textbf{Free-space THz time-domain spectroscopy system combined with the Split-Florida Helix Magnet~\cite{curtis2018broadband}}. The broadband THz pulses are generated by mixing the fundamental and its second-harmonic laser field (generated from a frequency-doubling BBO crystal) in a nitrogen-purged atmosphere. Laser pulses are focused and collimated using off-axis parabolic (OAP) mirrors. The THz wavevector is perpendicular to the applied magnetic field (Voigt geometry). For detection, THz pulses are focused onto a THz-air breakdown coherent detector (THz-ABCD). A portion of the fundamental beam is used to gate the THz-ABCD to recover the full electric field waveform.}
    \label{fig:helix}
\end{figure}

\begin{figure*}
    \centering
    \includegraphics[width=\textwidth]{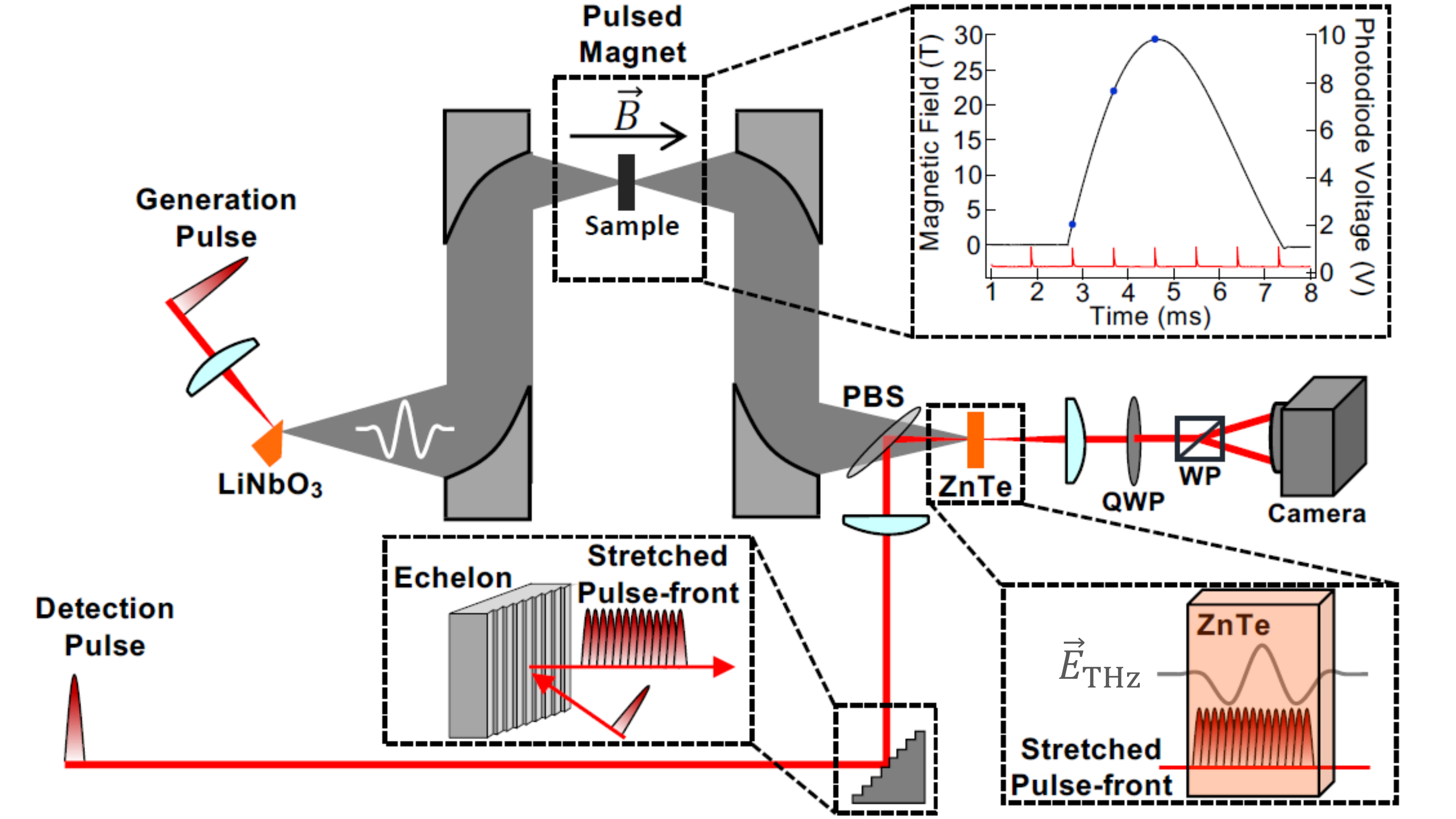}
    \caption{\textbf{Free-space single-shot THz time-domain spectroscopy system combined with a pulsed magnet~\cite{NoeetAl16OE,MakiharaetAl20arXiv}}. THz pulses strong enough for single-shot detection are generated using LiNbO$_3$. Detection is performed in a single-shot manner using a reflective echelon, which stretches the probe pulse in the time domain. The inset at the top shows that for a 1-kHz repetition-rate laser, three measurements can be done on the rising edge of the magnetic field pulse. A pellicle beamsplitter (PBS) is used to combine the THz and near-infrared probe beams. A quarter-wave plate (QWP) and a Wollaston prism (WP) are commonly used to detect the polarization rotation of the probe beam. }
    \label{fig:pulsed_magnet}
\end{figure*}

One of the earliest demonstrations of THz-TDS in strong magnetic fields combined a superconducting magnet with delay stage-based THz electric field sampling. This was achieved at the National High Magnetic Field Laboratory in Los Alamos, where a fiber-coupled photoconductive THz emitter and receiver were placed directly in the cryogenic bore of an 18-T superconducting magnet~\cite{crooker2002fiber}. The THz emitter and receiver were coupled by optical fibers to a Ti:Sapphire laser. Given the sensitivity of the emitted THz amplitude and spectrum on the inputted ultrafast optical pulse's temporal width, extreme attention was paid to dispersion of the optical gate pulse in the fiber. Specifically, to account for positive group velocity dispersion within the pulse, the inputted optical pulse was negatively chirped prior to entering the optical fiber. Further, generating optimum THz electric fields required precise positioning of the optical fiber on the THz emitter to within 1~$\mu$m. Thus, optical fiber-coupled THz emitters and receivers that were robust against thermal cycling were developed to enable temperature-dependent experiments down to $1.5$~K.

Given that fiber-coupled, photoconductive THz emitters and detectors can be challenging due to their sensitivity to dispersion of ultrashort laser pulses, thermal cycling, and applied magnetic fields, free-space THz-TDS poses an attractive alternative. An early demonstration of free-space THz time-domain magnetospectroscopy at Rice University~\cite{WangetAl07OL} combined a 10-T superconducting magnet with delay stage-based free-space THz-TDS using ZnTe crystals. The robustness of this system has enabled a wealth of studies of condensed matter phenomena occurring in the THz frequency range in magnetic fields up to 10-T \cite{WangetAl10NP,ArikawaetAl11PRB,ArikawaetAl12OE,ZhangetAl14PRL,ZhangetAl16NP,LietAl18NP,LietAl18Science}. A schematic of this system is illustrated in Fig.\,\ref{fig:SC_magnet}. The output from an amplified Ti:Sapphire laser is divided into: (1)~a generation beam that pumps a ZnTe crystal to generate THz radiation, and (2)~a detection beam that samples the THz electric field at different time delays using a delay stage. The THz radiation is guided through the superconducting magnet using parabolic mirrors and is overlapped with the time-delayed detection pulse using a pellicle beamsplitter (PBS). Free-space EO sampling using ZnTe is implemented to sample the THz electric field. The detection crystal is followed by a quarter-wave plate (QWP) and a Wollaston prism (WP) to implement balanced detection of the detection pulse's polarization~\cite{wu1996ultrafast}.

\subsection{DC Resistive Magnets}

Stronger static magnetic fields than those available in superconducting magnets can be achieved in resistive magnets~\cite{toth2011fabrication}. As suggested by its name, resistive magnets exploit a coil made of a resistive material. This not only demands large amounts of electrical power, but also requires advanced cooling mechanisms to force large amounts of water through the coil. As a result, high-field resistive magnets are often found at national laboratories.

THz TDS in a resistive magnet has been demonstrated in static magnetic fields up to 25~T in the recently developed Split Florida-Helix magnet~\cite{curtis2018broadband} at the National High Magnetic Field Laboratory in Tallahassee. This magnet was specifically designed to allow easy optical access to samples through four large, elliptic, interchangeable windows, making it ideal for THz time-domain magnetospectroscopy~\cite{curtis2014ultrafast,curtis2018broadband}. The minimum experimentally accessible temperature is 5~K. The magnetic field is generated perpendicular to the optical access, making it well-designed for experiments in the Voigt geometry. To access the Faraday geometry, custom sample holders can be used to steer the beam inside the magnet to be parallel to the field~\cite{paul2019coherent}. Recently, this magnet has been used to perform optical-pump THz-probe studies of GaAs quantum wells in magnetic fields up to 25~T~\cite{curtis2018broadband}. A schematic of the Split Florida-Helix system is illustrated in Fig.\,\ref{fig:helix}. THz radiation is generated using air-plasma generation~\cite{KimetAl08NP,kress2004terahertz}, guided through the magnet using parabolic mirrors, and detected using THz-ABCD and a delay stage to obtain temporal resolution.

\subsection{Pulsed Magnets}\label{pulsed_magnets_section}
Individuals seeking the strongest magnetic fields must forgo static magnetic fields for time-dependent, pulsed magnetic fields. Although stronger magnetic fields can be attractive for elucidating novel physics, the combination of pulsed magnetic fields and THz-TDS is challenging. Namely, as discussed in Section~\ref{Sampling Section}, one cannot easily sample THz waveforms using a delay stage in time-dependent magnetic fields. Thus, more sophisticated forms of sampling, also discussed in Section~\ref{Sampling Section}, must be implemented. Nonetheless, the allure of marrying magnetic fields exceeding 30~T with THz-TDS has motivated the development of several pulsed THz time-domain magnetospectroscopy facilities~\cite{MolteretAl10OE,MolteretAl12OE,NoeetAl13RSI,NoeetAl14AO,NoeetAl16OE,spencer2016terahertz,post2020observation,MakiharaetAl20arXiv}.

\begin{table*}
\caption{Summary of magnets for high-field THz time-domain magnetospectroscopy}
\label{table}
\centering
\begin{tabular}{ |p{2.7cm}||p{4cm}|p{4cm}|p{4cm}|}
 \hline
 \hline
  & Superconducting & Resistive & Pulsed\\
 \hline
 Magnetic Field & $\lesssim$ 10 T & $\lesssim$ 25 T & $\lesssim$ 30 T\\
 \hline
 THz Sampling Mechanism   & Delay Stage \cite{walecki1993terahertz,SomeNurmikko94APL,SomeNurmikko94PRB,SomeNurmikko96PRB,Crooker02RSI,WangetAl07OL,sumikura2007development,ikebe2008characterization,ScalarietAl12Science,george2012terahertz,wood2013chip,WuetAl16Science} & Delay Stage \cite{curtis2018broadband} & ECOPS~\cite{NoeetAl14AO,post2020observation}, ASOPS~\cite{spencer2016terahertz}, Reflective Echelons~\cite{NoeetAl16OE,MakiharaetAl20arXiv}, Rotating Delay Line~\cite{MolteretAl10OE}\\
 \hline
 Additional Notes & Superconducting magnets are easiest to implement and benefit from having static fields, albeit the weakest fields. & Resistive magnets provide the strongest static fields but are only available at national laboratories due to their demand for resources. & Pulsed magnets provide the strongest magnetic fields but require sophisticated protocols for THz sampling.\\
 \hline
\end{tabular}
\end{table*}

One facility that combines pulsed magnetic fields up to 30~T with THz-TDS was developed at Rice University~\cite{NoeetAl13RSI}. This magnet, known as the Rice Advanced Magnet with Broadband Optics (RAMBO), is a unique, table-top magnet that allows easy optical access to samples subjected to pulsed magnetic fields up to 30~T. The magnet measures only 17~cm in width with a bore diameter of 12~mm. This remarkable optical addressability makes RAMBO ideal for THz spectroscopy, yet the challenge of temporally sampling a THz electric in a time-dependent magnetic field remains. This challenge has been circumvented using ECOPS~\cite{NoeetAl14AO,post2020observation} and single-shot detection using reflective echelons~\cite{NoeetAl16OE}. A similar magnet combined with ASOPS has been developed at the University of Manchester~\cite{spencer2016terahertz}. A schematic of the RAMBO system is illustrated in Fig.\,\ref{fig:pulsed_magnet} when combined with echelon-based single-shot detection. The output from an amplified Ti:Sapphire laser is divided into: (1)~a generation pulse that generates THz radiation using the tilted-pulse-front excitation method in LiNbO$_3$ \cite{HeblingetAl08JOSAB}, and (2)~a detection pulse that samples the entire THz waveform by implementing single-shot detection. Specifically, the detection pulse is reflected off of a reflective echelon~\cite{MinamietAl13APL} that stretches the detection pulse front, allowing it to overlap with the entire THz waveform in the ZnTe detection crystal. The THz radiation is guided through the pulsed magnet using parabolic mirrors and is overlapped with the stretched detection pulse front using a PBS. Free-space EO sampling using ZnTe is implemented to sample the THz electric field. The THz electric field in the time domain can be extracted by imaging the reflective echelon surface through the ZnTe crystal, a QWP, and a WP~\cite{NoeetAl16OE,MakiharaetAl20arXiv}. The magnetic field profile (black line), output from the Ti:Sapphire laser (red line), and sampled magnetic field strengths (blue dots) are shown in the inset, where the output from the Ti:Sapphire laser is measured by a photodiode.

Another facility that combines pulsed magnetic fields up to 40 T with THz-TDS was developed at the Laboratoire National des Champs Magn\'{e}tiques Intenses in Toulouse, France~\cite{MolteretAl10OE}. In this facility, THz time-domain magnetospectroscopy measurements are taken in the reflection geometry in a magnet with a bore diameter of 30~mm. To rapidly sample the THz waveform, the authors use a rotating delay line, described in Section~\ref{Sampling Section}.

Table~\ref{table} shows a summary of different types of magnets combined with THz-TDS. For static magnetic fields up to 25~T produced by superconducting and resistive magnets, THz waveform sampling utilizing a standard delay stage can be easily employed. For higher magnetic fields, pulsed magnets have to be used, which necessitates one to implement ASOPS, ECOPS, rotating delay line, or single-shot (echelon) methods for THz sampling. To acquire a complete THz waveform (tens of ps), all aforementioned sampling techniques, except the single-shot method, require a data acquisition time of about 150~$\mu$s, which is short enough for magnetic field pulses on the order of a few ms. However, if one tries to increase the magnetic field strength by reducing the magnetic field pulse width to a $\mu$s scale, then the only possible option is the single-shot (echelon) THz sampling scheme.

\section{Physical Phenomena in High Magnetic Fields Probed by THz-TDS}\label{highlights}

A number of electronic, magnetic, and vibronic transitions in condensed matter systems occur in the THz frequency or meV photon energy range~\cite{McCombeWagner75Review1,NussOrenstein98THz,Mittleman03Book,Schmuttenmaer04CR,Lee09Book,BasovetAl11RMP,UlbrichtetAl11RMP,Dexheimer17Book}. An applied magnetic field can change these transitions in terms of frequency, width, and strength, and a systematic study of such changes can provide significant insight into the nature of the transitions. Furthermore, a high magnetic field can induce new phenomena, including field-induced phase transitions, and THz spectroscopy can serve as a sensitive probe of the quantum states, interactions, and dynamics of the system exhibiting such phenomena.  Specifically, the THz optical conductivity (or dielectric) tensor of a material as a function of frequency, temperature, and magnetic field often contains critical information on the microscopic physics behind complex many-body behaviors of solids. 

While Fourier-transform infrared spectroscopy in the THz range has been extensively used in conjunction with high magnetic fields to study low-energy phenomena in condensed matter systems, such as spin excitations~\cite{kezsmarki2014one, bordacs2012chirality, penc2012spin, peedu2019spin,TalbayevetAl08PRL2,mihaly2004field,mihaly2006spin,kezsmarki2015optical}, magnetoplasmons~\cite{autore2015observation}, quantum spin liquids~\cite{wang2017magnetic, sahasrabudhe2020high}, topological insulators~\cite{LaForgeetAl10PRB,schafgans2012landau}, and superconductors~\cite{schafgans2010towards}, the focus of this section is on THz-TDS experiments performed in high magnetic fields. For the purposes of this review, we consider high magnetic fields to be those higher than 5~T.     

\subsection{Cyclotron Resonance}

A free electron of mass $m^*$ in a magnetic field, $B$, undergoes a spiral motion with angular frequency $\omega_\text{c} = eB/m^*$, the cyclotron frequency.  When electromagnetic radiation with angular frequency $\omega$ interacts with the spiraling electron, resonant absorption, i.e., cyclotron resonance (CR), occurs if $\omega = \omega_\text{c}$.   CR spectroscopy has long been used for characterizing the band structure of semiconductor materials~\cite{DresselhausetAl55PR,LaxMavroides60SSP,McCombeWagner75Review1,McCombeWagner75Review2,Kono01MMR,KonoMiura06HMF,HiltonetAl12Book}. Since $B$ and $\omega$ can be known with high precision, $\omega_\text{c}$ and hence $m^*$ can be determined precisely. In addition, the resonance linewidth of CR contains information on scattering properties of carriers.  Within the Drude model, the quality ($Q$) factor of a CR line is given by $\omega_\text{c}\tau$, where $\tau$ is the scattering time, which can be precisely determined once $\omega_\text{c}$ is determined.  Furthermore, the mobility $\mu = e\tau/m^*$ can be deduced from the knowledge of $m^*$ and $\tau$.

\begin{figure}[hbtp]
    \centering
    \includegraphics[width=0.49\textwidth]{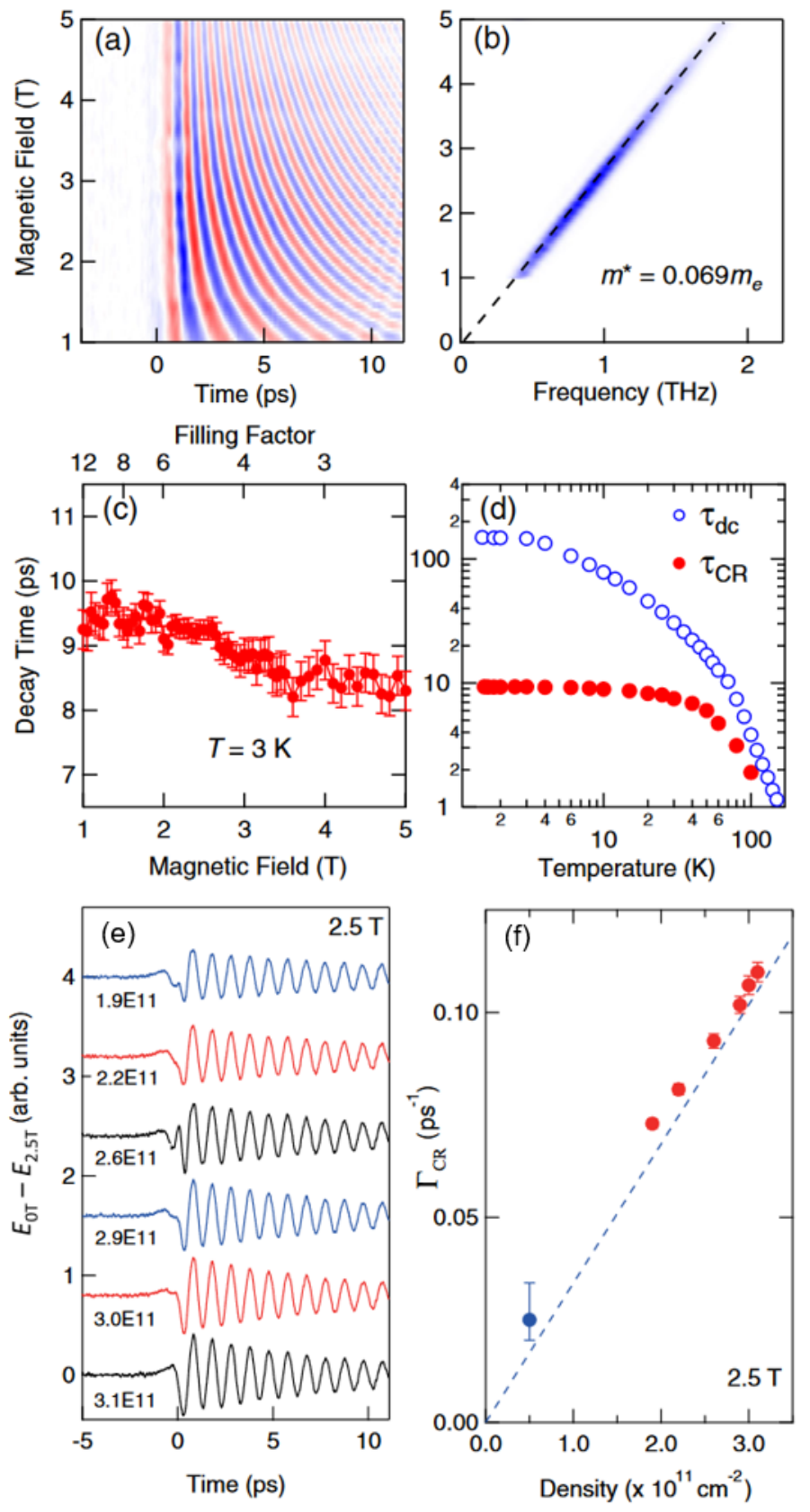}
    \caption{\textbf{Superradiant decay of cyclotron resonance in ultrahigh-mobility two-dimensional electron gases}~\cite{ZhangetAl14PRL}. (a)~Magnetic field dependence of CR oscillations in the time domain, showing  peaks  (blue)  and  valleys  (red). (b)~The frequency-domain version of (a). Black dashed line: a linear fit with a cyclotron mass of $0.069 m_0$. (c)~Magnetic field dependence of the CR decay time, $\tau_\text{CR}$ at 3~K. (d)~Temperature dependence of $\tau_\text{CR}$ at 2.5~T and the dc scattering time, $\tau_\text{dc}$, determined from the DC mobility, $\mu = e\tau_\text{dc}/m^*$. (e)~CR oscillations in the time domain for samples with different densities by controlling the illumination time. (f)~CR decay rate as a function of electron density. Blue solid circle: sample 2 (low density). Red solid circles: sample 1 (high density). The blue dashed line represents a theoretical relation expected from the phenomenon of superradiance~\cite{Dicke54PR} without any adjustable parameter (with $n_\text{GaAs}=3.6$ and $m^*=0.069 m_0$).}
    \label{fig:CR_QiPRL2014}
\end{figure}

A unique ability of THz-TDS is to obtain dynamical information directly from time-domain data, as well as to perform quantum control experiments.  The CR coherence time can be accurately determined through the decay of the CR amplitude as a function of time~\cite{WangetAl10OE}, and multiple THz pulses can be used to manipulate the coherent dynamics of a Landau-quantized 2D electron gas (2DEG)~\cite{ArikawaetAl11PRB}. Furthermore, Zhang \textit{et al}.~\cite{ZhangetAl14PRL}\ demonstrated that in an ultrahigh-mobility 2DEG the CR coherence time is determined by a \textit{superradiant} decay, i.e., cooperatively accelerated radiative decay~\cite{Dicke54PR}. The authors performed systematic studies on the CR decoherence of 2DEGs using THz-TDS in magnetic fields as a function of temperature, magnetic field, electron density, and mobility. Most importantly, it was demonstrated that the CR decay time is inversely proportional to the density; see Fig.\,\ref{fig:CR_QiPRL2014}.  Experimental results were explained by a fully quantum mechanical theory, identifying superradiant damping as the dominant decay mechanism of CR at low temperatures~\cite{ZhangetAl14PRL}.

The main advantage of using high magnetic fields in CR experiments is that CR becomes observable in large-mass, low-mobility materials~\cite{KonoMiura06HMF}. CR can be observed only when the $Q$-factor $\omega_\text{c}\tau = B_\text{r} \mu$ is larger than 1, where $B_\text{r}$ is the resonance magnetic field. This condition is not easy to fulfill when $m^*$ is large or $\tau$ is small.  Unfortunately, many strongly correlated materials have heavy masses, and newly discovered materials typically have short scattering times, requiring high magnetic fields to observe CR~\cite{MiuraetAl91SSC,KonoetAl93Physica,KonoetAl93PRB1,KonoetAl93PRB2,spencer2016terahertz, knap1997cyclotron, wang1996magneto,post2020observation}. 

Recently, in a topological crystalline insulator, Pb$_{0.5}$Sn$_{0.5}$Te, THz conductivity measurements identified two types of CR~\cite{cheng2019magnetoterahertz}. The authors found different magnetic field dependencies of the cyclotron frequency and scattering rate for the two types of carriers, which they attributed to different nontrivial topological properties of their band structures. This work demonstrates that magnetoterahertz response can be used to isolate signatures of bulk states in Dirac and Weyl semimetals and study their nontrivial band geometry and Berry curvature that can be extracted from direct non-contact CR and field-dependent scattering rate measurements~\cite{cheng2019magnetoterahertz}.

\subsection{Magnetoexcitons}

Excitons are bound states of an electron and a hole. Due to the long-range electron-electron, hole-hole, and electron-hole (e-h) Coulomb interactions, systems of e-h pairs can exhibit a variety of possible phases depending on the pair density, temperature, and magnetic field~\cite{Jeffries75Science}. If an applied perpendicular magnetic field is sufficiently strong, exciton-exciton interactions can vanish in an ideal 2D system due to an e-h charge symmetry (hidden symmetry). This results in cancellation of Coulomb interactions between excitons and, therefore, formation of ultrastable 2D magnetoexcitons~\cite{MacDonaldRezayi90PRB,DzyubenkoLozovik91JPA,ApalkonRashba91JETP,RashbaetAl00SSC}. 

Zhang \textit{et al}.\ have investigated the stability of an insulating exciton gas against a Mott transition~\cite{ZhangetAl16PRL}. The Mott transition is driven by density-dependent Coulomb interactions from an insulating excitonic gas into a metallic e-h plasma. However, the presence of the magnetic field protects the excitons from forming a plasma through ionization. The stability of excitons was revealed by performing optical-pump/THz-probe measurements to observe intraexciton transitions in a 10-T direct superconducting magnet with 150\,fs optical pulses and a THz bandwidth of 2.5\,THz. As the pump intensity increased, the excitons exhibited a clear Mott transition at 0~T, while the $1s-2p_-$ intraexciton resonance persisted at the highest pump intensity at 9~T.  Interestingly, however, 2D magnetoexcitons did dissociate under thermal excitation (a large magnetic field but higher temperatures), even if they are stable against a Mott transition~\cite{ZhangetAl16PRL}.

More recently, Li \textit{et al}.~\cite{LietAl20arXiv} reported an observation of THz gain in 2D magnetoexcitons in GaAs quantum wells in magnetic fields up to 8~T. The gain had a strict selection rule: it was observable only for hole-CR-active circular polarized THz radiation. The gain feature, represented by a negative optical conductivity, was a narrow-band sharp peak with a linewidth of $< 0.45$~meV, and its frequency shifted with the applied magnetic field. As the origin of this surprising phenomenon, the authors proposed the scattering of two free excitons into one biexciton in the presence of a magnetic field~\cite{LietAl20arXiv}.

\subsection{Faraday and Kerr Rotations}

Faraday and magneto-optical Kerr spectroscopies are powerful techniques  for investigating materials in which time-reversal symmetry is lifted, either by an internal or external magnetic field. The absence of time-reversal symmetry makes the off-diagonal elements of the optical conductivity (or dielectric) tensor finite, which leads to Faraday/Kerr rotations.  With the advent of polarization-dependent THz-TDS, it has become possible to study Faraday and Kerr effects in the THz range~\cite{hangyo2005terahertz}. 

A 2DEG in a strong perpendicular magnetic field naturally exhibits THz Faraday/Kerr rotations due to the finite Hall resistivity. At high enough magnetic fields and low enough temperatures, the quantum Hall effect (QHE) appears~\cite{KlitzingetAl80PRL}, with zero longitudinal resistivity and quantized Hall resistivity. Recently, Ikebe \textit{et al}.~\cite{IkebeetAl10PRL} observed an \textit{optical} Hall effect in a 2DEG at THz frequencies, showing that the Faraday rotation angle is of the order of the fine-structure constant in a quantum Hall plateau region~\cite{IkebeetAl10PRL}. A plateau behavior, a hallmark of the QHE, in the Faraday angle/optical conductivity versus magnetic field was seen at the filling factor $\nu=2$, for magnetic fields between $4.3$~T$<B<5.4$~T at 3~K. As the temperature increased to $20$~K, the classical limit was achieved. 

\begin{figure}[hbtp]
    \centering
    \includegraphics[width=0.45\textwidth]{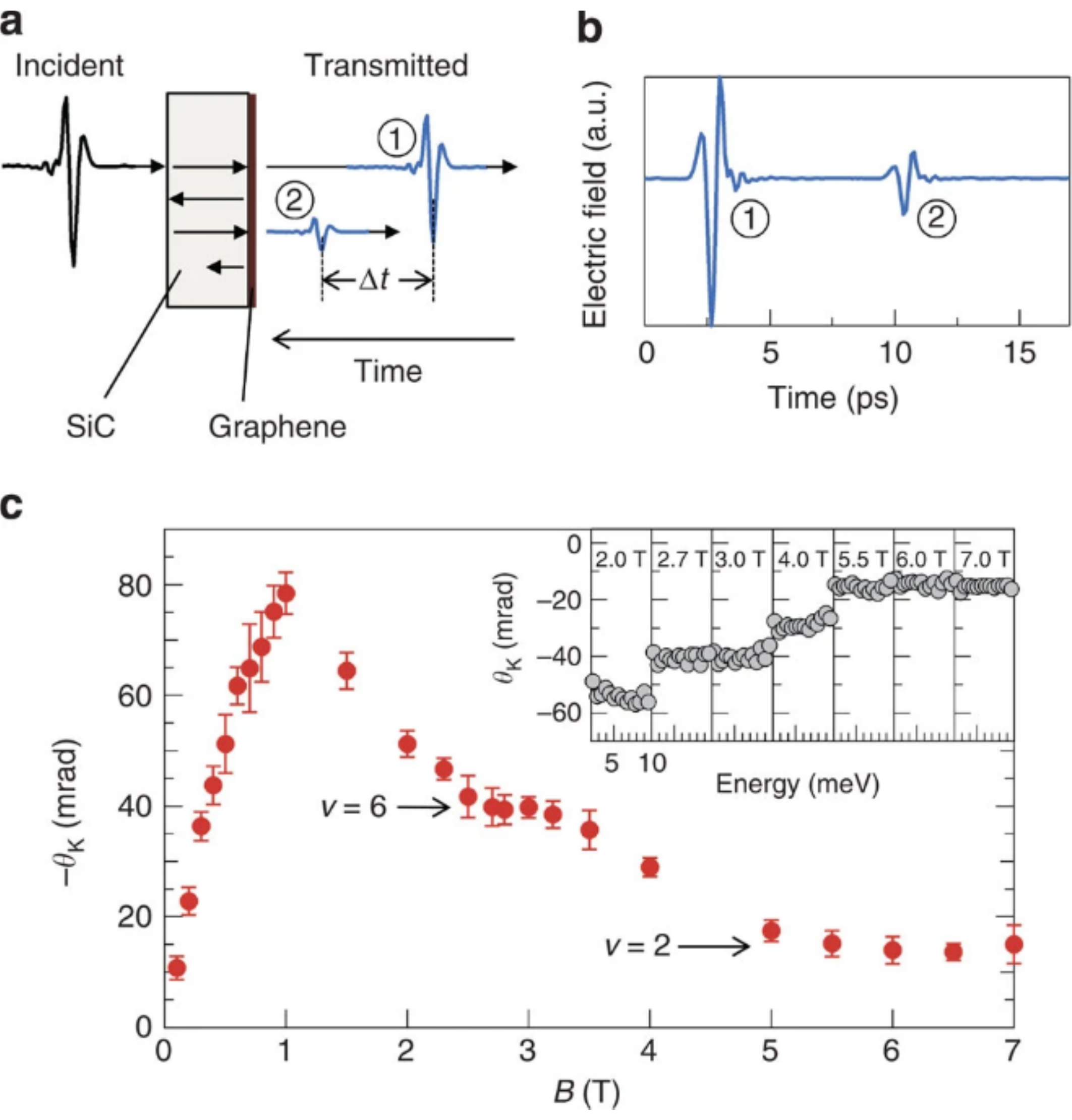}
    \caption{\textbf{Magneto-optical Kerr rotation in monolayer graphene}. (a)~Schematic of the experimental configuration depicting the transmission of a THz pulse through the sample. (b)~The time-domain waveform of the THz pulse transmitted through the sample. (c)~The magnetic field dependence of the Kerr rotation at 1~THz for the sample with a Fermi energy of 70~meV. The inset shows Kerr rotation spectra at the indicated magnetic fields. Adapted from Ref.~\cite{ShimanoetAl13NC}}
    \label{fig:graphene_Kerr}
\end{figure}

Shimano \textit{et al}.~\cite{ShimanoetAl13NC} have measured THz Faraday and Kerr rotations in monolayer graphene in the QHE regime, where both rotations exhibit values defined by the fine-structure constant. Figure~\ref{fig:graphene_Kerr}a shows schematic of the propagation of a THz pulse through the sample, which results in the appearance of two pulses in the time domain (see Fig.\,\ref{fig:graphene_Kerr}b). The polarization state of the first pulse determines the Faraday angle, while the second pulse contains information on both Faraday and Kerr rotations. Therefore, the Kerr angle can be obtained in a transmission geometry along with the Faraday angle. The magnetic field dependence of the Kerr angle at 1~THz is shown in Fig.\,\ref{fig:graphene_Kerr}c, where the $\nu=2$ state is reached above 5.5~T. Both Faraday (not shown) and Kerr rotations defined by the fine-structure constant appear precisely at the quantum Hall steps expected for the Dirac electrons in graphene~\cite{ShimanoetAl13NC}. 

In general, any coupling between magnetic and electric properties of a material is called the magnetoelectric effect (ME)~\cite{Fiebig2005}. Recent combination of high magnetic fields and THz-TDS applied to the ferrimagnetic pyroelectric CaBaCo$_4$O$_7$ provided deeper insights into the microscopic exchange striction and its magnetoelectricity~\cite{yu2017terahertz}. The topological magnetoelectric effect (TME) refers to 3D topological insulators that can be characterized by bulk magnetoelectrics rather than surface conductors~\cite{Armitage2019, EssinetAl09PRL, MaciejkoetAl10PRL, MorimotoetAl15PRB, QietAl08PRB, TseMacDonald10PRL, TseMacDonald11PRB, WangetAl15PRB, ZhangetAl19PRL}. When a magnetic field is applied to a TI, it breaks time-reversal symmetry, which results in a gap in the surface Dirac state(s). Macroscopically, both the Hall conductivity of the sample surface and the bulk magnetoelectric response become quantized. In addition, this magnetoelectric coupling leads to the modified Maxwell's equations inside the topological insulator, analogous to the theory of axion electrodynamics in particle physics~\cite{Wilczek87PRL}.

\begin{figure*}[hbtp]
    \centering
    \includegraphics[width=0.8\textwidth]{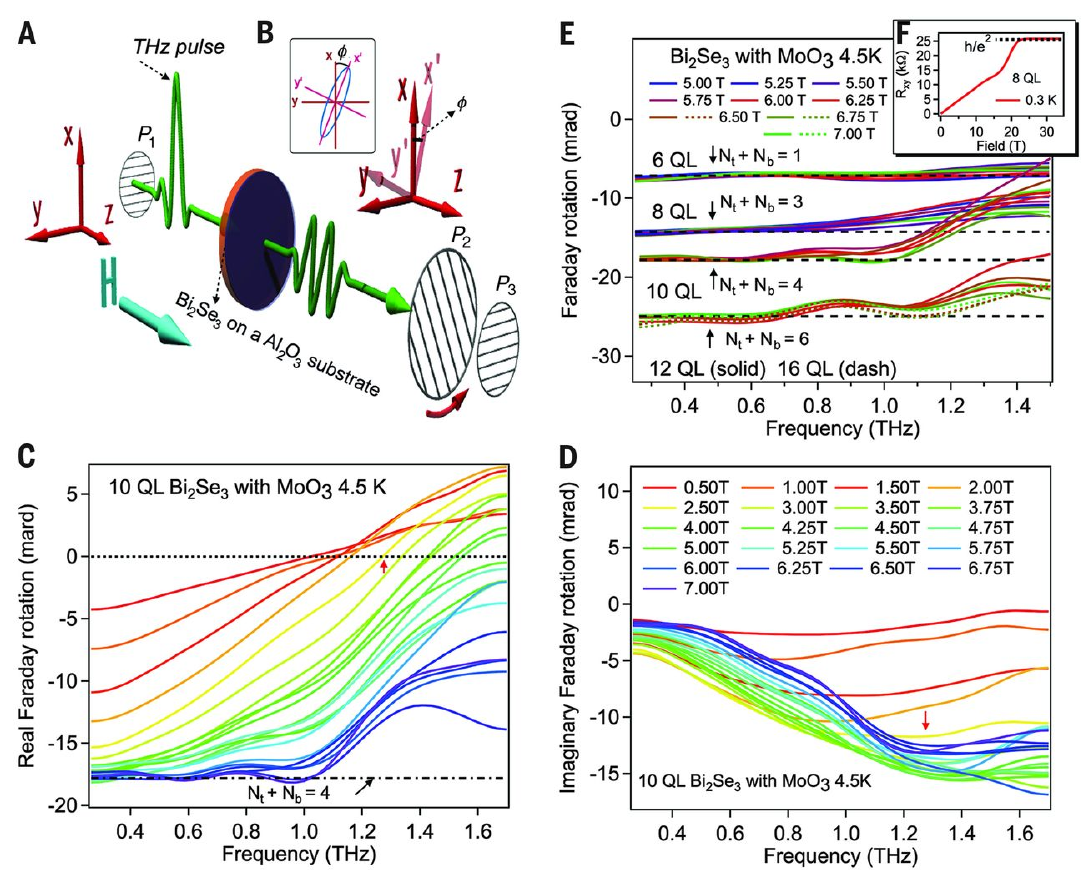}
    \caption{\textbf{Quantized Faraday rotation of topological surface states in Bi$_2$Se$_3$}. (A)~Schematic diagram of the Faraday rotation experimental setup. P1, P2, and P3 are polarizers. The polarization acquires an ellipticity simultaneously, as shown in (B). (C)~Real part of the Faraday rotation of 10-QL Bi$_2$Se$_3$ films with MoO$_3$ at 4.5~K at various magnetic fields [color-coded as in (D)]. The dash-dotted line is the theoretical expectation. (D)~Imaginary part of the Faraday rotation. A representative cyclotron frequency is marked by a red arrow for data at 2.5~T. (E)~Quantized Faraday rotation for different samples. Dashed black lines are theoretical expectation values assuming certain values for the filling factor of the surface states. (F)~DC Hall resistance of a representative 8-QL sample. Adapted from Ref.\,\cite{WuetAl16Science}}
    \label{fig:Wu2016_1}
\end{figure*}

The TME effect can be measured via Faraday and Kerr rotations at low photon energies, such as the microwave and far-infrared spectral ranges~\cite{TseMacDonald10PRL,TseMacDonald11PRB,MaciejkoetAl10PRL}. Several THz magneto-optical spectroscopy  experiments to probe the TME have been performed at small~\cite{HancocketAl11PRL} and moderate~\cite{Jenkins2010, LaForgeetAl10PRB, ValdesAguilaretAl12PRL, WuetAl15PRL, WuetAl16Science, DziometAl17NC} magnetic fields. Wu \textit{et al}.\ have used THz-TDS and observed quantization of Faraday and Kerr rotations at fields $>5$~T in Bi$_2$Se$_3$~\cite{WuetAl16Science}. Figure~\ref{fig:Wu2016_1}A shows a schematic diagram of the Faraday rotation experimental setup. The real and imaginary parts of the Faraday rotation are shown in Figs.\,\ref{fig:Wu2016_1}C and D, indicating that for the 10-QL sample (QL: quintuple layer, 1 QL $\approx$ 1\,nm) of Bi$_2$Se$_3$ above 5~T the low-frequency tail becomes flat and overlaps with higher field data at plateau values predicted by the theory. Similarly, the 6-, 8-, 12- and 16-QL samples enter the quantized regime as the low-frequency tails of real part of the Faraday rotation falls on their expected values (see Fig.\,\ref{fig:Wu2016_1}E). 
Figure~\ref{fig:Wu2016_1}F further shows that this effect is not the conventional DC QHE as the quantum Hall resistivity exhibits a plateaus in these films only above $\sim$24~T. 

More recently, Li \textit{et al}.\ reported THz Faraday and Kerr rotations in thin films of Bi$_{1-x}$Sb$_x$ in $B$ up to 30~T~\cite{LietAl19PRB}. Several samples with different compositions $x$ were studied. As $x$ increases, the Bi$_{1-x}$Sb$_x$ alloy shows a semimetal-to-topological-insulator transition. The Faraday and Kerr rotation measurements were conducted using a single-shot time-domain THz spectroscopy setup combined with a table-top pulsed minicoil magnet with $B$ up to 30~T, as described in Section~\ref{pulsed_magnets_section}. Faraday and Kerr rotation spectra for the semimetallic films ($x<0.07$) showed a pronounced dip that blue shifted with $B$, whereas spectra for the topological insulator films ($x>0.07$) were positive and featureless. Ellipticity spectra for the semimetallic films showed resonances, whereas the topological insulator films showed no detectable ellipticity. The evolution of Faraday and Kerr angles with $B$ for a fixed THz frequency of 0.7~THz showed a saturation behavior. It  might  be  tempting  to  explain it as the  quantized  optical Hall  effect  observed in  several  recent  studies~\cite{WuetAl16Science, DziometAl17NC, OkadaetAl16NC}. However, it turns out that this saturation behavior observed within $<30$~T can be explained as a summation of the broadened CR  signals  contributed  by  the  two  electron  pockets. Theoretical calculations predicted a major quantum Hall plateau in  the  60~T $< B <$ 110~T range. Overall, the combination of the THz-TDS polarimetry experiments performed in high $B$ and the detailed theoretical  analysis can be applied to other topological materials to investigate surface and bulk carrier contributions to the optical conductivity. 

\subsection{Spin Excitations}

Recent years have seen the development of an exciting area of research, within the general area of spintronics, which focuses on antiferromagnets (AFMs) instead of ferromagnets/ferrimagnets. In AFMs, magnonic excitations typically occur in the THz frequency range and can couple with another degree of freedom coming from the same magnetic system rather than excitations supplied by an external source.  
These THz magnons are promising for high-speed, low-power computing applications. THz-TDS is well suited for exciting and probing such THz magnons in AFMs materials. Applying a strong magnetic field makes the excitation frequencies even higher and can lead to an observation of novel phenomena. 

Morris \textit{et al}.~\cite{morris2014hierarchy} investigated kink bound states in 1D ferromagnetic Ising chain compound CoNb$_2$O$_6$. In addition to recently observed kink excitations by neutron scattering, high signal-to-noise ratio and high energy-resolution THz-TDS led to the appearance of new kink bound states. Magnetic field dependent measurements in the Faraday geometry were performed in order to understand the origin of these kink excitations, which were assigned to two- and four-kink excitations. An additional new feature observed at an energy somewhat less than twice the lowest bound state was interpreted as a novel bound state~\cite{morris2014hierarchy}.

Magnons in quantum spin liquid (QSL) candidate materials have also been studied using THz-TDS~\cite{little2017antiferromagnetic,wu2018field,ozel2019magnetic}. $\alpha-$RuCl$_3$ is a promising Kitaev QSL candidate, but it orders magnetically at low temperatures despite the Kitaev exchange interactions present in the effective spin Hamiltonian. However, the magnetic order can be suppressed by an in-plane magnetic field of about 7.5~T, where several THz-TDS experiments demonstrated signatures of a field-induced QSL state, but complete understanding of the observed excitations is still lacking~\cite{little2017antiferromagnetic,wang2017magnetic,shi2018field}.  Wu \textit{et al}.\ have investigated the field evolution of magnons in $\alpha-$RuCl$_3$. The susceptibility associated with the AFM state was measured as a function of applied DC field $H$ and THz probe field $B_\text{THz}$. Two sharp resonances and other broader features appearing only above approximately 4~T were attributed to zero-wave-vector magnons and a continuum of two-magnon excitations, respectively~\cite{wu2018field}. Subsequently, Ozel \textit{et al}.\ showed that the magnon behavior in an applied magnetic field can be understood only by including an off-diagonal exchange term to the minimal Heisenberg-Kitaev model~\cite{ozel2019magnetic}.

Another interesting example is the optical diode effect that appears in the paramagnetic phase of polar FeZnMo$_3$O$_8$~\cite{yu2018high}. The authors demonstrated 100-fold difference for light traveling in different opposite directions. Notably, this effect exists at high temperatures (up to 110~K) where FeZnMo$_3$O$_8$ is in the paramagnetic state and no long range magnetic order exists. The optical diode is explained to be happening at the electron spin  resonance frequency, which the authors assigned to the single-ion anisotropy gap excitation~\cite{yu2018high}.

\subsection{Ultrastrong Light-Matter Coupling}

Cavity quantum electrodynamics (QED) in the ultrastrong coupling (USC) regime has been recently studied using condensed matter systems~\cite{Forn-DiazetAl19RMP,KockumetAl19NRP}. In this unusual regime, realized due to the colossal dipole moments available in solids, as opposed to atoms, the normalized coupling strength $\eta = g/\omega_0$ becomes comparable to unity.  Here, $g$ is the light-matter coupling rate (or constant) and $\omega_0$ is the resonance frequency. Particularly, THz-TDS in magnetic fields is useful for studying phenomena in the USC regime because of the typical frequencies of CR. The continuous tunability of the matter frequency (i.e., the cyclotron frequency, $\omega_\text{c}$) with respect to a cavity photon mode frequency makes this system distinctly advantageous compared to other intraband transitions such as quantum-well intersubband transitions.

\begin{figure}[btp]
    \centering
    \includegraphics[width=0.45\textwidth]{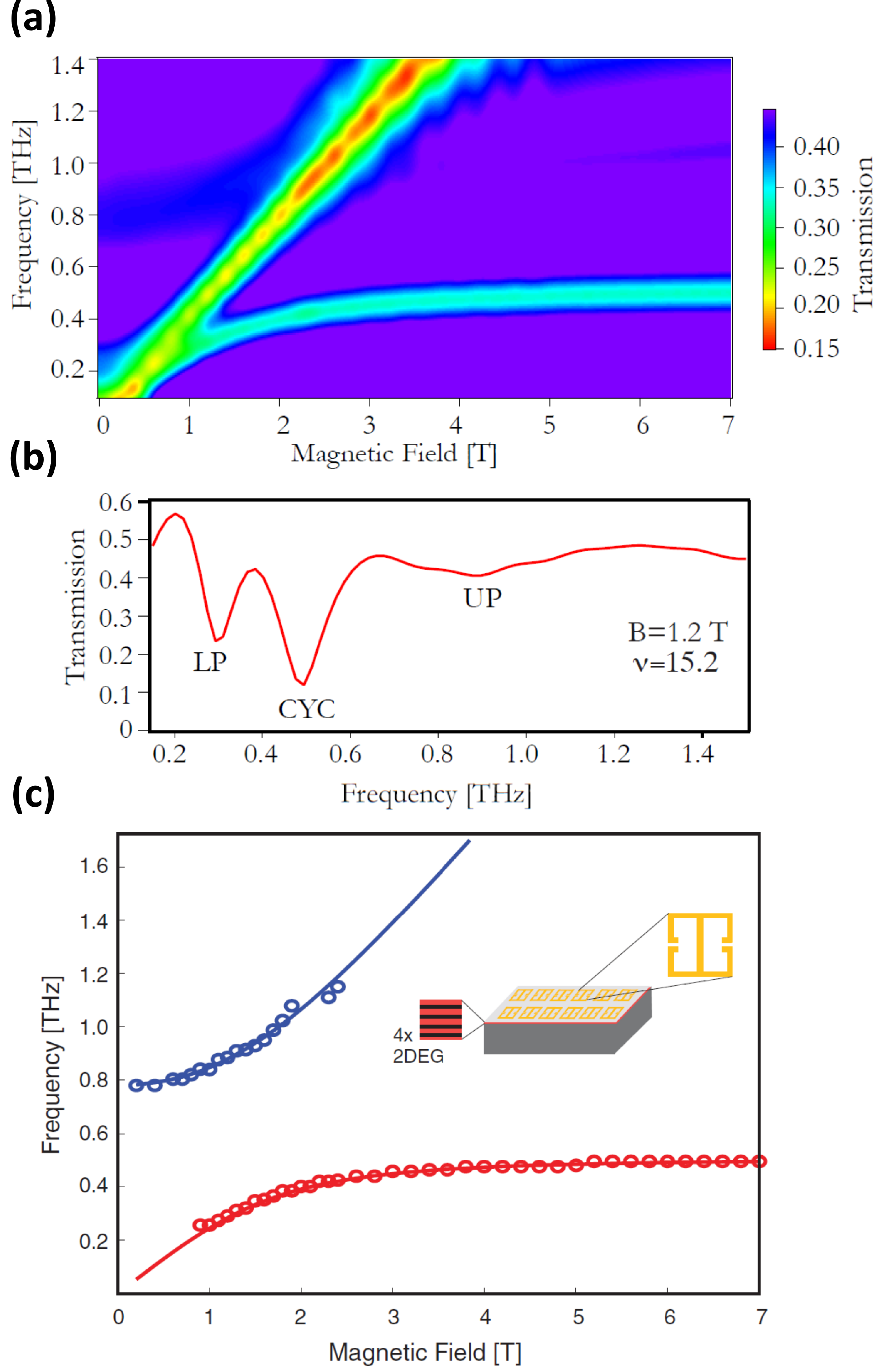}
    \caption{\textbf{Ultrastrong coupling between THz split-ring resonators and the CR of a high-mobility 2DEG.} (a)~Transmission as a function of $B$ at 10~K. (b)~A vertical cut from panel (a) in the anticrossing region at $B = 1.2$~T. Both the upper-polariton and lower-polariton modes are observed together with the free-space CR (CYC) arising from the uncovered regions of the 2DEG. (c)~Polariton branches plotted as a function of magnetic field. Solid lines are theoretical calculations. Inset: schematic of the 500~GHz resonator deposited on the sample surface. Adapted from Ref.~\cite{ScalarietAl12Science}.}
    \label{fig:Nicolas_Scalari}
\end{figure}

Scalari \textit{et al}.~\cite{ScalarietAl12Science} studied the USC between the CR of a high-mobility 2DEG and an array of metallic split-ring resonators (SRRs). Transmission measurements were carried out using a broadband THz-TDS system coupled to a cryostat equipped with a split-coil superconducting magnet. As observed in Fig.\,\ref{fig:Nicolas_Scalari}a, an anticrossing behavior between the first SRR mode and the CR of the 2DEG was observed. Since some regions of the 2DEG were not covered by the resonators, the uncoupled CR signal (CYC) is visible in the spectra together with the upper and lower polaritons (UP and LP, respectively) (see Fig.\,\ref{fig:Nicolas_Scalari}b). By using a modified geometry and shifting the SRR frequency down to 500~GHz, the authors achieved $\eta = 0.58$, putting the system in the USC regime. Figure~\ref{fig:Nicolas_Scalari}c shows the transmission minima (open circles) plotted as a function of magnetic field and the two polariton branches (solid lines) were fit by a theoretical model by Hagenm\"{u}ller \textit{et al}.~\cite{HagenmulleretAl10PRB}.

\begin{figure}[btp]
    \centering
    \includegraphics[width=0.45\textwidth]{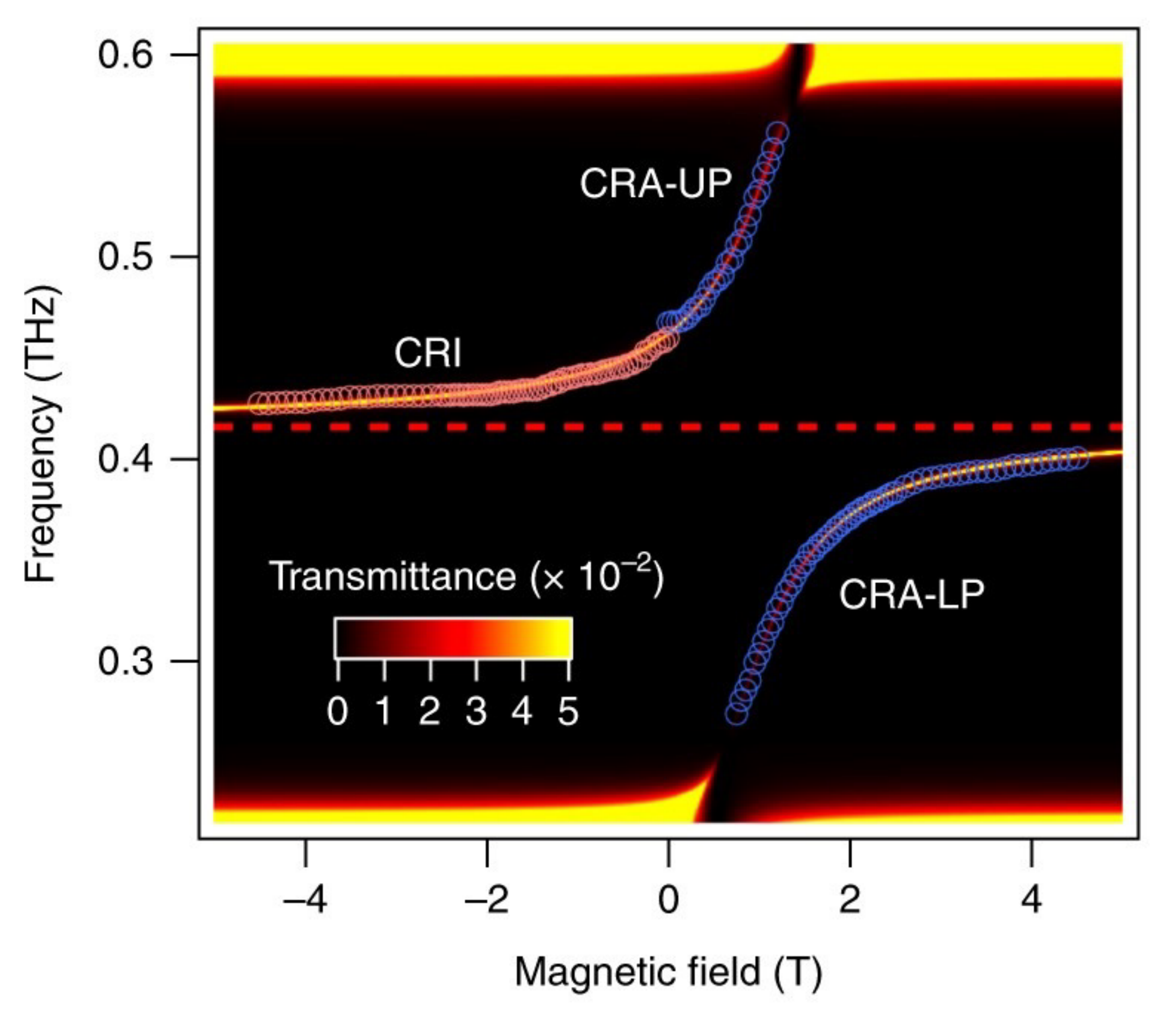}
    \caption{\textbf{Vacuum Bloch–Siegert shift in Landau polaritons with ultrahigh cooperativity~\cite{LietAl18NP}.} Transmission spectra in the vicinity of the anticrossing region between CR and the first cavity mode (red dashed line) for simulations using the full Hamiltonian (solid lines) is shown. The counter-rotating and $A^2$ terms have to be included in the full Hamiltonian. Experimental data points are shown as open circles.}
    \label{fig:Nicolas_Fig1}
\end{figure}

Li \textit{et al}.\ reported the observation of a vacuum Bloch-Siegert shift (VBSS) in CR of a 2DEG in a THz cavity QED setting~\cite{LietAl18NP}. The authors used a circularly polarized THz probe pulse to isolate the contributions coming from the counter-rotating terms in the Hamiltonian, therefore providing an unambiguous and direct demonstration of the VBSS. THz-TDS measurements were performed in the Faraday geometry using an achromatic THz quarter-wave plate. As shown in Fig.\,\ref{fig:Nicolas_Fig1}, three branches were resolved in the transmission spectra. Two of them, labeled CR-active (CRA) upper polariton and lower polariton (CRA-UP and CRA-LP, respectively), arise in the $B > 0$ region from the co-rotating interaction between the CR of electrons and the right-hand circularly polarized (RCP) probe beam. In the $B < 0$ region, the polarization of the probe is preserved but the negative $B$ value reverses the electronic CR trajectory, resulting in the observation of the CR-inactive (CRI) mode. Since the VBSS arises from the USC between matter and the counter-rotating vacuum field in the cavity, the authors found that by isolating and individually measuring the CRI mode it was possible to distinguish between the diamagnetic (or $A^2$) shift and the VBSS.

USC has also been observed in an unusual spin-magnon-coupled system~\cite{LietAl18Science}. Li \textit{et al}.\ observed USC between the electron paramagnetic resonance (EPR) of Er$^{3+}$ spins and the quasi-ferromagnetic (q$_{\text{FM}}$) mode of the Fe$^{3+}$ magnons in the rare-earth orthoferrite, Er$_{x}$Y$_{1-x}$FeO$_{3}$. An applied external DC magnetic field tuned the EPR frequency with respect to the fixed q$_{\text{FM}}$ magnon mode frequency, and a clear anticrossing behavior was resolved in the THz range. Moreover, a $g \propto \sqrt{N}$ behavior (so-called Dicke cooperativity~\cite{Dicke54PR}), where $N$ is the net density of EPR-contributing Er$^{3+}$ spins, was obtained. The authors estimated a maximum $\eta = 0.18$ in one of the configurations, achieving USC for this matter-matter coupled system.

\begin{figure*}
    \centering
    \includegraphics[width = 0.8\linewidth]{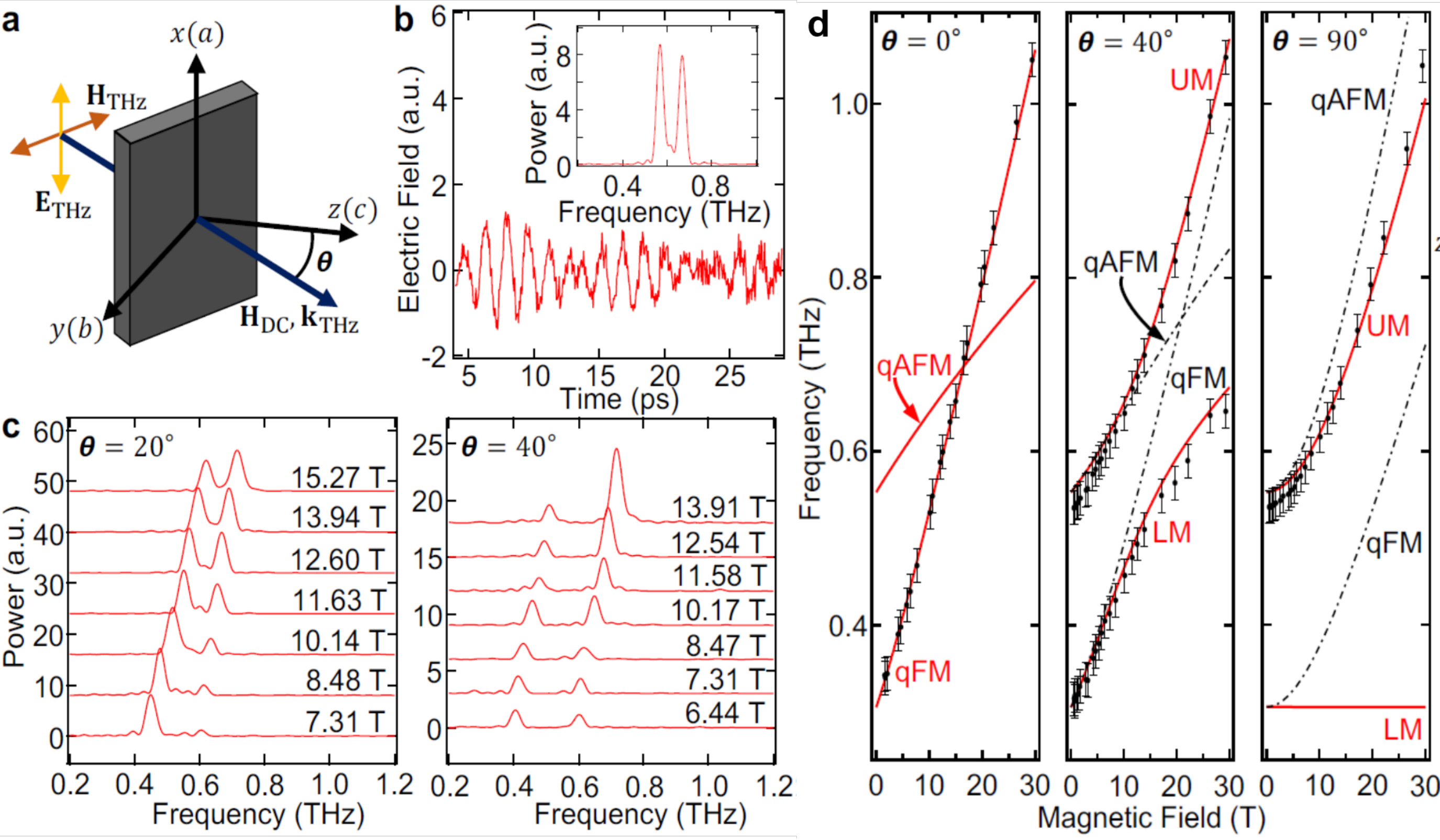}
    \caption{\textbf{THz-TDS of magnon-magnon ultrastrong coupling in  YFeO$_3$ evidencing dominant vacuum Bloch-Siegert shifts~\cite{MakiharaetAl20arXiv}}. (a)~Schematic of THz magnetospectroscopy studies of YFeO$_3$ in a tilted magnetic field. $\mathbf{H}_\text{DC}$ was applied in the $b-c$ plane at an angle of $\theta$ with respect to the $c$-axis, with $\mathbf{k}_\text{THz}//\mathbf{H}_\text{DC}$ and $\mathbf{H}_\text{THz}$ polarized in the $b-c$ plane. (b)~Transmitted THz waveform for $\theta = 20^{\circ}$ at $\mu_0|\mathbf{H}_\text{DC}| = 12.60$~T displaying beating in the time domain and two peaks in the frequency domain. (c)~Magnon power spectra for $\theta = 20^{\circ}$ and $\theta = 40^{\circ}$ at different $\mathbf{H}_\text{DC}$, displaying larger frequency splitting for larger $\theta$. (d)~Experimentally measured magnon frequencies for $\theta = 0^{\circ}, 40^{\circ}, 90^{\circ}$ versus $\mu_0|\mathbf{H}_{\mathrm{DC}}|$ (black dots) with calculated resonance magnon frequencies (solid red lines) and decoupled qFM and qAFM magnon frequencies (black dashed-dotted lines). The upper mode frequency becomes lower than the qAFM frequency at $\theta = 90^{\circ}$, indicating a dominant vacuum Bloch-Siegert shift compared to the vacuum Rabi splitting-induced  shift.}
    \label{fig:YFeO3}
\end{figure*}

Most recently, an observation has been made of USC between the qAFM and qFM magnon modes in YFeO$_3$~\cite{MakiharaetAl20arXiv}. The authors demonstrated that this novel matter-matter USC simulates a unique cavity QED Hamiltonian with tunable coupling strengths and dominant counter-rotating interactions. Figure \textcolor{blue}{\ref{fig:YFeO3}}a shows a schematic of the experiment. The YFeO$_3$ crystals were cut such that the external magnetic field $\mathbf{H}_\text{DC}$ could be applied in the $b-c$ plane at different angles $\theta$ with respect to the $c$-axis. An example of a transmitted THz waveform for $\theta = 20^{\circ}$ at $\mu_0|\mathbf{H}_\text{DC}| = 12.60$~T is shown in Fig. \textcolor{blue}{\ref{fig:YFeO3}}b demonstrating beating in the time domain and, consequently, two peaks in the frequency domain. Fig. \textcolor{blue}{\ref{fig:YFeO3}}c demonstrates the magnetic field dependence of these two magnon peaks for $\theta = 20^\circ$ and $\theta = 40^\circ$, where frequency splitting is clearly larger for $\theta = 40^\circ$. Figure \textcolor{blue}{\ref{fig:YFeO3}}d plots experimentally measured magnon frequencies (black dots) for $\theta = 0^{\circ}, 40^{\circ}, 90^{\circ}$ as a function of applied magnetic field, $\mu_0|\mathbf{H}_{\mathrm{DC}}|$. Theoretically calculated resonance magnon frequencies are plotted as red solid lines, demonstrating excellent agreement. Therefore, the authors found that the applied magnetic field could be used to tune the vacuum Rabi splitting and the VBSS. Further, in certain geometries, the frequency shifts of the coupled modes were dominated by the VBSSs and not the vacuum Rabi splitting-induced shifts. This is particularly clear when $\theta = 90^\circ$, where the red shift from the VBSS dominates the blue shift from the vacuum Rabi splitting-induced shift, leading to the upper mode being lower in frequency than the decoupled qAFM mode. A well-established microscopic spin model of this material system~\cite{Herrmann63JPCS} accurately reproduced their observed resonances without any adjustable parameters. Moreover, it was shown that this lightless spin model can be precisely mapped to a polariton model described by an \textit{anisotropic} Hopfield Hamiltonian in which the magnon-magnon coupling strengths are easily tunable and the counter-rotating terms dominate the co-rotating terms, consistent with their observation of giant VBSSs. Finally, they theoretically showed that the ground state is intrinsically squeezed, consisting of a two-mode squeezed vacuum as expected in the USC regime~\cite{Artoni-Birman89QO,Schwendimann-Quattropani92EPL,CiutietAl05PRB}, with quantum fluctuation suppression as large as $5.9$~dB. 


\section{Summary and Outlook}

In conclusion, we reviewed the current state-of-the-art of the rapidly advancing techniques for combining strong magnetic fields with THz time-domain spectroscopy.  We described different types of high-field magnets, including superconducting, resistive, and pulsed magnets, and various THz radiation detection schemes. Table~I gives a summary of maximum field strengths for each magnet type and provides a short guide to a reader looking to build a similar experimental setup. These experimental developments have already led to observations of new exciting physical phenomena, some of which were also reviewed in this article, including superradiant decay of cyclotron resonance, stability of 2D magnetoexcitons against a Mott transition, quantized Faraday and Kerr rotations, coherent THz magnon excitations, and ultrastrong light-matter and matter-matter coupling in the THz range. Clearly, the combination of time-domain THz spectroscopy with high magnetic fields is a fertile field of research in condensed matter physics, many exciting new discoveries are expected to be made, and combining even higher magnetic fields with THz spectroscopy will allow researchers to investigate a broader range of materials. 

\section*{Acknowledgments}

J.K.\ acknowledges support from the U.S.\ Army Research Office (W911NF-17-1-0259), the U.S.\ National Science Foundation (NSF MRSEC DMR-1720595), the U.S.\ Department of Energy (DEFG02-06ER46308), and the Robert A.\ Welch Foundation (C-1509).

\bibliography{biblio/jun_merged}

\end{document}